\documentclass[conference]{IEEEtran}
\IEEEoverridecommandlockouts
\usepackage{cite}
\usepackage[letterpaper,top=0.8in,bottom=1.15in,left=0.75in,right=0.75in]{geometry}
\usepackage{amssymb,amsfonts}
\usepackage{graphicx}
\usepackage{textcomp}
\usepackage{overpic}
\usepackage{verbatim} 
\usepackage{pifont}
\usepackage{times,amsmath,color,amssymb,graphicx, epstopdf, epsfig,cite,psfrag,subfigure,algorithm}
\usepackage[noend]{algpseudocode}
\usepackage{enumerate}

\usepackage{xcolor}
\def\BibTeX{{\rm B\kern-.05em{\sc i\kern-.025em b}\kern-.08em
    T\kern-.1667em\lower.7ex\hbox{E}\kern-.125emX}}
\begin{document}

\title{ 
    


    Pattern-Aware Virtual Network Embedding Optimization for Cloud Data Centers

}

	

\author{\IEEEauthorblockN{Binquan Guo$^{a,c}$, Zhou Zhang$^b$, Junfeng Zhai$^a$, Zheng Zhang$^a$, Marie Siew$^c$, and Zehui Xiong$^d$}\\

    	\IEEEauthorblockA{
        $^a$School of Telecommunications Engineering, Xidian University, Xi'an, P. R. China\\
        $^b$College of Computer Science and Electronic Engineering, Hunan University, Changsha, P. R. China\\
        $^c$Pillar of ISTD, Singapore University of Technology and Design, 487372, Singapore\\
        $^d$School of Electronics, Electrical Engineering and Computer Science, Queen's University Belfast, BT9 5BN Belfast, U.K.\\
Email:bqguo@stu.xidian.edu.cn, zt.sy1986@163.com, jfzhai@stu.xidian.edu.cn, zhangzheng@xidian.edu.cn, \\marie\_siew@sutd.edu.sg, z.xiong@qub.ac.uk
}
\thanks{This work was supported by A*STAR under its IAF-ICP (H25-MCP3438). 
The corresponding authors are Zhou Zhang and Marie Siew.
}

}


\maketitle

	\begin{abstract}

	The network virtualization (NV) technology has enabled the sharing of multiple resources among virtual networks (VNs) in cloud data centers. One of the key challenges is to allocate resources in real-time for virtual network request (VNR), which is known as online virtual network embedding (VNE). However, the existing online VNE methods do not exploit the multi-dimensional complementary relationship among diverse VNRs, resulting in the fragmentation and waste of substrate resources. In this paper, we propose the pattern matching based online VNE approach by constructing appropriate matching rules among observed patterns to maximize resources utilization. We devise the clustering based VNRs quantization method and conduct rigorous study on the pattern combination filtering problem by modeling it as an integer linear programming problem. Then, we utilize the column generation to solve it and construct the pattern matching rules. Based on the rules, we propose an online pattern matching VNE algorithm with linear worst-case complexity. Evaluation on a 106-server testbed using Alibaba production cluster trace dataset shows that our algorithm achieves close-to-offline performance and more accepted workloads that outperforms traditional designs by	25\%-30\%.
	
\end{abstract}



\section{Introduction}

As a promising technology, network virtualization (NV) has attracted significant interest in academia and industry and is widely used in data center networks. It enables multiple heterogeneous virtual networks (VNs) to share a common substrate network (SN), addressing the rigidity of traditional architectures. Each VN specifies its demands in a virtual network request (VNR), including computing resources for nodes and bandwidth for links. Efficient real-time mapping of VNRs onto the SN, known as online virtual network embedding (VNE) \cite{fischer2013virtual}, is essential. However, the unpredictable arrivals and diverse topologies of VNRs complicate embedding decisions and require effective admission control.

Recently, VNE has attracted significant research attention. It has been shown that VNE with joint node and link constraints can be reduced to the multi-way separator problem, which is NP-hard \cite{dolati2019deepvine}. Existing approaches are generally classified into exact and heuristic methods \cite{fischer2013virtual}. Exact methods target offline scenarios and solve MIP or ILP formulations with high computational complexity \cite{melo2013optimal}. For online scenarios, most heuristic methods adopt two-stage node and link mapping strategies, which reduce complexity but may cause resource fragmentation and low utilization \cite{chowdhury2011vineyard}. Hybrid approaches combine heuristic and exact methods to balance complexity and performance \cite{cao2017novel}. However, these methods often ignore historical information and VNR characteristics, leading to inefficient resource allocation.

Theoretically, traditional methods \cite{dolati2019deepvine, cao2017novel} mainly focus on the matching between VNRs and resources. In fact, inspired by \cite{lee1985simple}, VNE can be transformed to a pattern matching problem, so that the online algorithm can approximate the offline performance. However, the matching method in \cite{lee1985simple} is designed for single dimension resource and cannot be directly applied to multi-dimensional scenarios. 
The reason is that the complementary rules in multiple dimensions cannot be formulated without the pattern characteristics of VNRs, where the historical data is required. Authors in \cite{naori2020online} considered the characteristics of historical data. However, their objective is to maximize the acceptance rate, which cannot guarantee the resource utilization of substrate nodes.
Moreover, the authors in \cite{thakkar2020muvine} aimed at maximizing the utilization of physical resources through reinforcement learning. However, the prediction based model suffers from unbalanced resource allocation and cannot widely adapt to the production data center environment.


In this paper, we propose a pattern matching based online VNE framework that maximizes resource utilization through appropriate matching rules. 
We devise the clustering based VNR quantization method and model the pattern combination filtering problem as an integer linear programming problem. Then, we utilize the column generation method to solve it and construct the pattern matching rules.
Based on the rules, we propose Online Pattern Matching VNE algorithm with worst case complexity of O($\mathcal{N}$).
Testbed evaluation on the Alibaba cluster trace shows near offline performance and 25\%-30\% improvement over traditional methods in accepted workloads.

\section{System Model and Problem Formulation}


\subsection{Substrate Network}


We consider a typical substrate network composed of a large number of servers in a data center, connected via a switch-centric topology. The network is modeled as an undirected weighted graph, where each node represents a server and each edge represents a link between servers. Resources are categorized into node and link types. Node resources include CPU, memory, and disk, while link resources refer to the bandwidth between servers.
The substrate network resource is denoted by $G=(\mathcal{N}, \mathcal{E}, \mathcal{A}_{\mathcal{N}},  \mathcal{A}_{\mathcal{E}})$, where

\textbullet \ $\mathcal{N} = \{\mathcal{N}_q | 1 \leq q \leq |\mathcal{N}|\}$ is the set of substrate nodes, and $\mathcal{N}_q$ denotes a server. 	

\textbullet \ $\mathcal{E}  = \{ (\mathcal{N}_u, \mathcal{N}_v) | \mathcal{N}_u \in \mathcal{N}, \mathcal{N}_v \in \mathcal{N}, u \neq v \}$ is the link set of the substrate network. Specifically, 
each pair of servers is connected, thus $|\mathcal{E}| = |\mathcal{N}| \cdot (|\mathcal{N}|-1)/2 $.

\textbullet \ $\mathcal{A}_{\mathcal{N}} = \{C(\mathcal{N}_{q})  | 1 \leq q \leq |\mathcal{N}|\} $ represents the node attribute of the substrate network, where $C(\mathcal{N}_{q}) = [c_q^1,c_q^2,...,c_q^d] $ is the capacity vector of node $\mathcal{N}_{q}$. Each element in $C(\mathcal{N}_{q})$ corresponds to the capacity of one type resource at node $\mathcal{N}_{q}$.

\textbullet \ $\mathcal{A}_{\mathcal{E}} = \{ b_{(u, v)} | (\mathcal{N}_u, \mathcal{N}_v) \in \mathcal{E}\} $ represents the link attribute of the substrate network. Specifically, $b_{(u, v)} $ is the weight value of bandwidth for the link $(\mathcal{N}_u, \mathcal{N}_v)$. It is assumed $b_{(u, v)} = b_{(v, u)}$, $\forall \mathcal{N}_u, \mathcal{N}_v \in \mathcal{N}, u \neq v$. Using the $b_{(u, v)}$, the bandwidth capacity of the node $\mathcal{N}_{q}$ denoted as $c_q^b$ can be calculated as
	\begin{equation*}
c_q^b = \max_{\forall \mathcal{N}_v \in \mathcal{N}} \{b_{(q,v)} \text{ or } b_{(v,q)} \}, q \neq v, \forall \mathcal{N}_q \in \mathcal{N}.
\end{equation*}
Furthermore, during the embedding process, the remaining resource vector at node $\mathcal{N}_{q}$ is denoted as $\mathcal{R}_q^c =  [r_q^1,r_q^2,...,r_q^d]$. Similarly, the remaining bandwidth resource at link $(\mathcal{N}_u, \mathcal{N}_v)$ is denoted as $r_{uv}^b$. Using the $r_{uv}^b$, the remaining bandwidth of the node $\mathcal{N}_{q}$ denoted as $r_q^b$ is calculated as
	\begin{equation*}
	r_q^b = \max_{\forall \mathcal{N}_v \in \mathcal{N}} \{r_{qv}^b \text{ or } r_{vq}^b \}, q \neq v, \forall \mathcal{N}_q \in \mathcal{N}.
	\end{equation*}
Besides,  we assume that servers are fully connected as in \cite{thakkar2020muvine}. For any two nodes $\mathcal{N}_u$ and $\mathcal{N}_v$, as long as their remaining bandwidth is nonzero, there exists at least one path between them with capacity $\min(r_u^b, r_v^b)$, which can be obtained using shortest path algorithms such as Dijkstra's algorithm.



\subsection{Virtual Network}
The users send a virtual network request (VNR) in the form of an undirected graph $G^v=\{\mathcal{N}^v, \mathcal{E}^v, \mathcal{A}_{\mathcal{N}^v}, \mathcal{A}_{\mathcal{E}^v}\}$. Each $G^v$ is comprised of a set of virtual machines (VMs) denoted as 
$\mathcal{N}^v = \{\mathcal{V}_i|1 \leq j \leq |\mathcal{N}^v|\}$ and a set of virtual links $\mathcal{E}^v=\{ (\mathcal{V}_i, \mathcal{V}_j)| \mathcal{V}_i \in \mathcal{N}^v, \mathcal{V}_j \in \mathcal{N}^v, i \neq j \}$.
Each VM $\mathcal{V}_i$ is associated with resource demand vector $\mathcal{D}_i^v \in \mathcal{A}_{\mathcal{N}^v}$, where $\mathcal{D}_i^v = [d_i^1,d_i^2,...,d_i^d]$.
 The bandwidth demand of the virtual link between VM $\mathcal{V}_i$ and $\mathcal{V}_j$ is denoted by $b_{ij}^v$. Then, the bandwidth demand of a VM $\mathcal{V}_i$ is calculated as
\begin{equation*}
d_i^b = \sum_{\forall \mathcal{V}_j \in \mathcal{N}^v} b_{ij}^v, i \neq j, \forall  \mathcal{V}_i \in \mathcal{N}^v .
\end{equation*}




\subsection{Problem formulation}
For each upcoming VNR, we define the binary variable  $x_{iq}$ to indicate if VM $\mathcal{V}_i$ is assigned to server $\mathcal{N}_q$. If $\mathcal{V}_i$ is assigned to $\mathcal{N}_q$, $x_{iq}=1$, and otherwise $0$. 
Further, for each VM, by merging its bandwidth demand $d_i^b$  into its resource demand vector $\mathcal{D}_i^v$, we can form an augmented demand vector $\tilde{\mathcal{D}}_i$, where $\tilde{\mathcal{D}}_i = [d_i^1,d_i^2,...,d_i^d, d_i^b]$.

Similarly, for each server $\mathcal{N}_{q}$, we combine its remaining resource vector $\mathcal{R}_q^c $ with its remaining bandwidth $r_q^b$ to form an augmented remaining vector $\tilde{\mathcal{R}}_q^c = [r_q^1,r_q^2,...,r_q^d, r_q^b]$. Its augmented capacity vector can be constructed by combining the resource vector $C(\mathcal{N}_q)$ with $c_q^b$ as $\tilde{C}(\mathcal{N}_q) = [c_q^1,c_q^2,...,c_q^d, c_q^b]$, which is $(d+1)$-dimensional like $\tilde{\mathcal{D}}_i$ and $\tilde{\mathcal{R}}_q^c$.

\textbf{Objective:} The objective of online VNE is to maximize the average resource utilization while satisfying resource capacity constraints. The objective function is written as
\begin{equation*}
\mathop \text{max }\limits_{x_{iq} \in \mathbf{x}}  U = \sum_{x \in \{1,d+1\}}
\frac{w_x \cdot \sum_{\mathcal{N}_q \in \mathcal{N}} {\frac{c_q^x - r_q^x}{c_q^x}} }
{\underbrace{|\{ q | \sum_{x \in \{1,d+1\}}{(c_q^x - r_q^x)} \neq 0, \forall \mathcal{N}_q \in \mathcal{N} \} |}_{\mathbf{E1}}},
\end{equation*}
where $U$ is the average resource utilization of the servers, and $w_x$ is the predefined weight factor of the $x$-th type resource such as CPU, memory, and bandwidth of each node. The symbol $| \cdot |$ is the cardinality of a set, and expression $\mathbf{E1}$ represents the number of servers that are currently used.

We consider that each type of different resources have the same weight, i.e., $w_x = \frac{1}{d+1}$, thus $U$ is calculated as
\begin{equation*}
U = \frac{\sum_{\mathcal{N}_q \in \mathcal{N}}  [{({\tilde{C}_q - \tilde{R}_q}) \odot \frac{1}{\tilde{C}_q} ] }}
{\lvert \{q| [({\tilde{C}_q - \tilde{R}_q}) \odot \mathbf{1}^{T}] \neq 0, \forall \mathcal{N}_q \in \mathcal{N} \} \rvert },
\end{equation*}
where $\mathbf{1} \in \mathbb{R}^{(d+1)\times 1}$, and 
$[({\tilde{C}_q - \tilde{R}_q}) \odot \mathbf{1}^{T}]$ is the sum of each type resource utilization of node $\mathcal{N}_q$. The symbol $\odot$ is the Hadamard product.

\textbf{Constraints:} For each upcoming VNR $G^v$, the following constraints must be satisfied.

\subsubsection{Place at most once}
Each VM can be assigned to only one machine as specified in (\ref{c_once_placement}), meaning no VM is split. However, this does not restrict how many VMs from the same VNR can be embedded on a single server.
\begin{equation}
\sum_{\mathcal{N}_q \in \mathcal{N}} {x}_{iq} <= 1, \forall \mathcal{V}_i \in \mathcal{N}^v. \label{c_once_placement}
\end{equation}
\subsubsection{Accept all VMs or reject}
A VNR must be mapped as a whole, meaning all its VM demands must be satisfied as in (\ref{c_map_as_whole}); otherwise, the entire VNR is rejected.
\begin{equation}
\sum_{\mathcal{N}_q \in \mathcal{N}} {x}_{iq} = \sum_{\mathcal{N}_q \in \mathcal{N}} {x}_{jq}, \forall \mathcal{V}_i, \mathcal{V}_j \in \mathcal{N}^v, i \neq j. \label{c_map_as_whole}
\end{equation} 

\subsubsection{Capacity constraints}
When a VM is placed on a server, the resources in all dimensions must be sufficient to meet the total demand of all VMs assigned to it, as in (\ref{c_capacity}).
\begin{equation}
\sum_{\mathcal{V}_i \in \mathcal{N}^v} d_i^x \cdot {x}_{iq} <= r_q^x, \forall \mathcal{N}_q \in \mathcal{N}, \forall 1 \leq x \leq d. \label{c_capacity}
\end{equation} 
\subsubsection{Mutual exclusion (optional)}
For security and reliability, VMs from the same VNR should be placed on different servers as in (\ref{c_mutual_exclusive}), which is optional.
\begin{equation}
\sum_{\mathcal{V}_i \in \mathcal{N}^v} {x}_{iq} \leq 1, \forall \mathcal{N}_q \in \mathcal{N} \label{c_mutual_exclusive}
\end{equation}

Finally, the online VNE problem can be formulated as 
\begin{equation*}
\begin{split}
\mathbf{OP: }\mathop \text{max }\limits_{x_{iq} \in \mathbf{x}} & U \\
\text{s.t. } & (\ref{c_once_placement})-(\ref{c_mutual_exclusive}).
\end{split}
\end{equation*}
The \textbf{OP} is an online multi-dimensional bin packing problem and is NP-hard. Even with optimal placement for each incoming VNR, long-term resource utilization remains suboptimal. In online settings, resource fragmentation often occurs when one resource type is exhausted while others remain unused, and load balancing cannot fully prevent this accumulation. Therefore, we aim to design an online approach that achieves near-offline performance by exploiting VNR patterns.



\section{The Pattern matching based VNE Approach}

In this section, we propose an online VNE method based on quantized VNR patterns that exploits multi-dimensional complementarity, consisting of two stages.




\subsection{Quantization-Based VNR Pattern Representation}

In order to overcome the complexity caused by the diversity of VNR, we quantify the VNRs to group similar VNRs together and divide them into disjoint sets.

\textcolor{black}{\textbf{Definition 1 } (VNR pattern).
	\textit{A VNR pattern refers to a complete set composed of a group of VMs from the real-world VNRs with similar resource demands, in which the Euclidean distance between the maximum and minimum elements is less than a specific parameter, and the demand of the VMs within it can be uniformly represented by the maximum element of the set with the allowable waste.}}
\begin{algorithm}
\caption{Clustering based Pattern Quantization}
	\label{array-sum}
	\hspace*{0.02in} {\bf Input:} Historical VN dataset 
	$\mathcal{H}=\{G^v_1,G^v_2,...,G^v_h\}$\\
	\hspace*{0.02in} {\bf Output:}  	Pattern quantization model $m*$, \\ \hspace*{0.54in}
	pattern maximum point representative set $\mathcal{P}$,\\ \hspace*{0.54in} pattern statistics $p(\mathcal{P})$.
	\begin{algorithmic}[1]
		\item Initialize $k^* = -1$,  $\mathcal{P} = \{\ \}$,  $p(\mathcal{P}) = \{\ \}$, $\mathcal{X} = \{\ \}$.
		\item  \textbf{for} each VN $G^v_l$ in VN dataset $\mathcal{H}$ \textbf{do} 
		\item  \hspace*{0.02in} \hspace*{0.02in}  \textbf{for} each VM $V_i$ in VN $G^v_l$ \textbf{do}
		\item  \hspace*{0.02in}  \hspace*{0.02in} \hspace*{0.02in} \hspace*{0.02in} $\mathcal{X} \longleftarrow \mathcal{X} \cup \{\mathbf{\tilde{\mathcal{D}}}_i\}$.
		\item  \textbf{for} each $k$ in  $[1,2, ..., k_{\emph{max}}]$ \textbf{do} (parallel)
		\item \hspace*{0.02in} \hspace*{0.02in}  Define a candidate model $\emph{m} = \emph{KMeans}(k)$.
		\item \hspace*{0.02in} \hspace*{0.02in}  Feed $\mathcal{X}$ for training the model by calling $\emph{m.fit}(\mathcal{X})$.
		\item \hspace*{0.02in} \hspace*{0.02in}  Classify $\mathcal{X}$ into $k$ categories $ \hat{\mathcal{X}}_h \subseteq \mathcal{X}, h \in [1,2,...,k]$.
		\item \hspace*{0.02in} \hspace*{0.02in}   \textbf{if }  $\forall  h \in [1,2,...,k]$, $||\emph{max}(\hat{\mathcal{X}}_h)-\emph{min}(\hat{\mathcal{X}}_h) ||_2 \leq \theta $  \textbf{then }
		\item  \hspace*{0.02in}  \hspace*{0.02in} \hspace*{0.02in} \hspace*{0.02in} $m^*  \longleftarrow  m, k^*  \longleftarrow  k$,  \textbf{break}.
		\item  \textbf{if }  $ k* \leq 0 $  \textbf{then }
		\item \hspace*{0.02in} \hspace*{0.02in}   \textbf{return } // No feasible model found, and stop procedure.
		\item \textbf{for} each $h \in [1,2,...,k^*]$ \textbf{do} 
		\item \hspace*{0.02in} \hspace*{0.02in}  $ \mathcal{P} \longleftarrow  \mathcal{P} \cup \{\emph{max}(\hat{\mathcal{X}}_h)\}$,  $ p(\mathcal{P}) \longleftarrow  p(\mathcal{P}) \cup \{ \frac{|\hat{\mathcal{X}}_h|}{|\mathcal{X}|} \times 100\% \}$
		\item  {\bf return}    $m^*$, $\mathcal{P},p(\mathcal{P})$.
	\end{algorithmic}
\end{algorithm}

In a given period, we have a set of historical VNRs $\mathcal{H}$. VNR patterns can be obtained by applying clustering models to $\mathcal{H}$. If the number of clusters is known in advance, the K-Means algorithm can be used to quantify both historical and newly arrived VNRs. In \textbf{Algorithm 1}, we present a pattern quantization procedure to obtain VNR patterns.


Firstly, assume the number of patterns does not exceed $k_{max}$. Each pattern can be denoted by $\hat{\mathcal{X}_{h}}, 1 \leq h \leq k_{max}$.
Firstly, we define a threshold $\theta$ to represent the Euclidean distance between the minimum element $min(\hat{\mathcal{X}_{h}})$ and the maximum element $max(\hat{\mathcal{X}_{h}})$ within each pattern.
By traversing over the historical data, the augmented vector of all VMs are collected into set $\mathcal{X}$ as the input of the clustering algorithm.
For each $1 \leq k \leq k_{max}$, we initialize the K-Means model and feed $\mathcal{X}$ for training the parameters. The value of $k$ will be adjusted repeatedly until the maximum Euclidean distance conditions $||\emph{max}(\hat{\mathcal{X}}_h)-\emph{min}(\hat{\mathcal{X}}_h) ||_2 \leq \theta, 1 \leq h \leq k $ are satisfied. If all the conditions are satisfied, the resulting quantization model $m*$ is obtained,  and the maximum element $max(\hat{\mathcal{X}_{h}})$ of each set $\hat{\mathcal{X}_{h}}$ is taken as the representative element to form a pattern set $\mathcal{P}$. Finally, the statistics of patterns are obtained and collected into a set $p(\mathcal{P}) = \{ p(\mathcal{P}_z) | \forall \mathcal{P}_z \in \mathcal{P}\}$.



\subsection{Statistics-Aware Pattern Combination \textcolor{black}{Filtering} Problem}
Based on the quantized VNR patterns, we further define the concept of pattern combination to represent a feasible packing scheme of substrate nodes.

\textbf{Definition 2 } (Pattern combination).
	\textit{A pattern combination is defined as a vector $\mathcal{B}_r$, where each element $ a_{hr} \in \mathcal{B}_r$ is a nonnegative integer representing the maximum number of the $h$-th pattern expected in a packing scheme $\mathcal{B}_r$, $h \in [1,2,...,k^*]$. During virtual network embedding, each server will be labeled with one pattern combination, e.g., $\mathcal{S}_q \leftarrow \mathcal{B}_r= (a_{1r},a_{2r},...,a_{hr},...,a_{k^*r})^T$,  indicating the server  $\mathcal{S}_q$ 
is expected to 
host at most $a_{hr}$ numbers of $h-$th pattern.}

Let $\Psi_x$ be the expected utilization threshold for $x$-type resource, where $\Psi_x \in (0,100\%], x \in [1,d+1]$.
We regard a pattern combination whose resource utilization upper bound in every dimensions are higher than the expected thresholds as a superior combination.
Obviously, one superior combination cannot fit all real-world VNR distributions.
Consider a scenario, in which the probability of pattern $\mathcal{P}_{less}$ is small, applying a superior combination expecting a large number of  $\mathcal{P}_{less}$, will inevitably lead to the idleness of servers, since pattern $\mathcal{P}_{less}$ may not occur for a long period.
Therefore, we need to design a method to filter the superior patterns combinations suitable for the given VNR statistics.

In order to obtain the superior pattern combinations for a given VNR statistics, we formulate the Statistics-Aware Pattern Combination \textcolor{black}{Filtering} Problem.
Firstly, let $\mathcal{B}$ be the set composed of all pattern combinations, where each $\mathcal{B}_r \in \mathcal{B} $ represents a pattern combination.
We define \textcolor{black}{an integer variable $\lambda_{\mathcal{B}_r}$}, where $\lambda_{\mathcal{B}_r} > 0$ means that the combination $\mathcal{B}_r$  will be selected as a matching rule, otherwise $\lambda_{\mathcal{B}_r} = 0$.
Moreover, we define a companion variable for each $\lambda_{\mathcal{B}_r}$ as $\tilde{\lambda}_{\mathcal{B}_r} \in \{0,1\}$, and the two variables are associated in the following way.
\begin{equation}
\lambda_{\mathcal{B}_r} -1 \leq M \cdot \tilde{\lambda}_{\mathcal{B}_r} - \epsilon \cdot (1-\tilde{\lambda}_{\mathcal{B}_r}). \label{c_companion_varible}
\end{equation}
In (\ref{c_companion_varible}), if $\lambda_{\mathcal{B}_r} > 0$, then $\tilde{\lambda}_{\mathcal{B}_r}=1$.
$M$ is a big positive constant.
Let $c^x \in \mathcal{C}$ be the capacity of $x-$type resource, $x \in [1, d+1]$.
The Statistics-Aware Pattern Combination Filtering Problem is formulated as
\begin{align}
\mathbf{MP: } \text{min} & \sum_{B_r \in \mathcal{B}}\lambda_{\mathcal{B}_r}, \nonumber \\
\text{s.t. }  & (\ref{c_companion_varible}), \nonumber
\\ & \sum_{\mathcal{B}_r \in \tilde{\mathcal{B}}} a_{hr} \lambda_{\mathcal{B}_r} \geq p(\mathcal{P}_h) M, \forall h \in [1,2,...,k^*], \label{c_demand_distribution}\\
&  \sum_{a_{hr} \in \mathcal{B}_r} a_{hr} \mathcal{P}_{h}^x  \geq \Psi_x \tilde{\lambda}_{\mathcal{B}_r} , \forall \mathcal{B}_r \in \mathcal{B},  \label{c_threshold_utilization}\\
&  \lambda_{\mathcal{B}_r} \in \mathbf{N}, \tilde{\lambda}_{\mathcal{B}_r} \in \{0,1\}, \forall \mathcal{B}_r \in \mathcal{B}, \label{c_integer_constraints}
\end{align}
where $\mathcal{P}_{h}^x $ is the $x$-th value in $\mathcal{P}_h$, $ x \in [1, d+1]$.  
Note that \textbf{MP} is in the form of \textbf{ILP}, and its aim is to solve a subset of pattern combinations with desired average resource utilization expectation, while covering all patterns of the given VNR statistics. 
When $|\mathcal{B}|$ is small, it can be solved by the brute-force search.
However, when $|\mathcal{B}|$ is large, searching for its optimal solution will be intractable. Therefore, we design a column generation based method to reduce its solving complexity.


\subsection{Column Generation-Based Pattern Combination Selection}


Intuitively, solving \textbf{MP} requires all pattern combinations in advance, which is intractable since the number of combinations $|\mathcal{B}|$ grows exponentially with the number of VNR patterns and resource types. To address this, column generation (CG) is used to iteratively generate better pattern combinations using only a small subset of variables in each iteration \cite{desrosiers2005primer}. Initially, a small set of feasible pattern combinations $\hat{\mathcal{B}}$ is used for \textbf{MP}, including at least one combination for each pattern $\mathcal{P}_h \in \mathcal{P}$. 
Then, we remove the constraints (\ref{c_threshold_utilization}) and define the restricted master problem (RMP) as
\begin{align}
\mathbf{RMP: } \text{min} & \sum_{B_r \in \hat{\mathcal{B}}}\lambda_{\mathcal{B}_r}, \nonumber \\
\text{s.t. } & \sum_{\mathcal{B}_r \in  \hat{\mathcal{B}}} a_{hr} \lambda_{\mathcal{B}_r} \geq p(\mathcal{P}_h) M, \forall h \in [1,2,...,k^*] \label{r_demand_distribution}\\
&   \lambda_{\mathcal{B}_r} \in \mathbf{N}, \forall \mathcal{B}_r \in  \hat{\mathcal{B}}.  \label{c_restricted_constraints}
\end{align}
\textbf{RMP} is initialized by a set of feasible pattern combinations $\hat{\mathcal{B}}$, and each pattern corresponds to at least one feasible pattern combination in $\hat{\mathcal{B}}$. In practice, we can form an initial $\hat{\mathcal{B}}$ by initializing $k^*$ pattern combinations. For instance, suppose each pattern combination contains only one pattern, e.g. $\mathcal{P}_h$, then the maximum feasible number $\lfloor \frac{\tilde{\mathcal{C}}}{\mathcal{P}_h} \rfloor $ of this pattern in a server can be calculated. By building a $k^*$-dimensional vector with  $h$-th value as $\lfloor \frac{\tilde{\mathcal{C}}}{\mathcal{P}_h} \rfloor $ and filling other positions with $0$, a feasible pattern combination for pattern $\mathcal{P}_h$ can be obtained.

Then, the linear programming relaxation of \textbf{RMP} is solved to obtain dual variables $\pi_h$ corresponding to constraints (\ref{c_demand_distribution}). 
\begin{align}
\mathbf{LRMP: } \text{min} & \sum_{B_r \in \hat{\mathcal{B}}}\lambda_{\mathcal{B}_r}  \nonumber \\
\text{s.t. } & (\ref{r_demand_distribution}) \nonumber \\
&  \lambda_{\mathcal{B}_r} \geq 0, \forall \mathcal{B}_r \in \hat{\mathcal{B}}.
\end{align}
The dual variables $\pi_h$ are provided to the pricing problem \textbf{PP} to generate new superior pattern combination. In detail,
\begin{align}
\mathbf{PP: }  \text{max} & \sum_{h = 1}^{k^*} \pi_h z_h  \nonumber \\
\text{s.t } & \sum_{h = 1}^{k^*} \mathcal{P}_h^x z_h \leq c^x, \label{c_capacity_constaints}\\
&   \sum_{h = 1}^{k^*} \mathcal{P}_h^x z_h \geq  \Psi_x c^x,  \\
&  z_h \geq 0, 1 \leq h \leq k^*, z_h \in \mathbf{N},
\end{align}
where integer variable $z_h$ denotes the number of pattern $\mathcal{P}_h$ included in a new pattern combination. The constraint (\ref{c_capacity_constaints}) ensures the new generated combination cannot exceed the total capacity of substrate resources among each dimension.
\textbf{PP} is designed to find feasible pattern combinations with 
the minimum reduced cost $\psi = 1- \sum_h^{k^*} \pi_h z_h $.
\begin{algorithm}[!t]	\caption{Column Generation Based Superior Pattern Combination Selection Algorithm}
	\label{array-sum}
	\hspace*{0.02in} {\bf Input:} Quantized pattern set $\mathcal{P}$, and its statistics $p(\mathcal{P})$. \\
	\hspace*{0.02in} {\bf Output:}
	Pattern combination set $\mathcal{S}$.
	\begin{algorithmic}[1]
		\item Initialize $A = [\frac{\tilde{\mathcal{C}}}{\mathcal{P}_1},...,\frac{\tilde{\mathcal{C}}}{\mathcal{P}_h},...,\frac{\tilde{\mathcal{C}}}{\mathcal{P}_{k^*}}  ]$, $\hat{\mathcal{B}} =\{ a_h | a_h \in {\lfloor diag(A) \rfloor}\}$.
		\item  {Solve} the \textbf{LRMP} with $\hat{\mathcal{B}}$ to obtain the dual variable $\mathbf{\pi}$ and put $\mathbf{\pi}$ into \textbf{PP}.
		\item  \textbf{while} $ \psi = 1- \sum_h \pi_h z_h  \leq \epsilon $ \textbf{do} 
		\item  \hspace*{0.02in} \hspace*{0.02in} 
		{Solve} \textbf{PP} by B\&C to obtain \textbf{new column}  $\mathcal{Z}_{\emph{opt}}$ and $ \psi$. 
		\item  \hspace*{0.02in} \hspace*{0.02in} {Add} new column to current \textbf{LRMP}(
		$\hat{\mathcal{B}} \longleftarrow \hat{\mathcal{B}} \cup  \{\mathcal{Z}_{\emph{opt}}\}$).
		\item  \hspace*{0.02in}  \hspace*{0.02in}  {\bf Go} to $5$.
		\item  { Solve} \textbf{MP} with $\hat{\mathcal{B}}$ using the B\&C to obtain optimal $\mathbf{\lambda}_{\hat{\mathcal{B}}}$.
		\item  { Set} $\mathcal{S} = \{\hat{\mathcal{B}}_x | \lambda_{\mathcal{B}_x} > 0, \forall \hat{\mathcal{B}}_x \in \hat{\mathcal{B}} \}$,
		\item  {\bf return}   $\mathcal{S}$.
	\end{algorithmic}
\end{algorithm}

\textbf{LRMP}s and \textbf{PP}s are solved iteratively until the termination condition is met. 
If $\psi < \epsilon$, the new generated pattern combination from \textbf{PP} is added to \textbf{LRMP}. 
If $\psi \geq \epsilon$, the new generated pattern combination from \textbf{PP} is no longer added to \textbf{LRMP}. Finally, to guarantee the optimal solution, \textbf{MP} is solved using existing columns, by replacing $\mathcal{B}$ with  $\hat{\mathcal{B}}$ and utilizing the Branch and Cut (B\&C) algorithm \cite{desrosiers2005primer,guo2022optimal}. 
The $\epsilon $ is a \textcolor{black}{negative} small constant near to zero, typically in the order of $\textcolor{black}{-}10^{-4}$, 
allowing numerical inaccuracies in the computed reduced costs \cite{desrosiers2005primer}. The detailed procedure is presented in \textbf{Algorithm 2}.
Note that $|\hat{\mathcal{B}}|$ can be much smaller than $|\mathcal{B}|$, which shortens the solution procedure.


\subsection{Online Pattern Matching VNE Algorithm} 

With the filtered pattern combinations $\mathcal{S}$, the Online Pattern Matching Algorithm is proposed to quickly embed each VNR. We regard each pattern combination $\mathcal{S}_x \in \mathcal{S}$ as a scheme \textcolor{black}{(or a tag)}, which is used to label each server during node mapping.
\begin{algorithm}[!t]	\caption{Online Pattern Matching VNE Algorithm}
	\label{array-sum}
	\hspace*{0.02in} {\bf Input:}
	Upcoming VNRs, SN $G=(\mathcal{N}, \mathcal{E})$\\
	\hspace*{0.02in} {\bf Output:}
	Embedding decision for each  $G^v\longrightarrow G$.
	\begin{algorithmic}[1]
		\item {\bf Initialization:}
		\item  Build a matrix $\mathbf{S} \in \mathbf{R}^{|\mathcal{S}|\times |\mathcal{P}|}$ from $\mathcal{S}$, where each column is a matching rule, and each row corresponds to a pattern.
		\item  Label a few servers with arbitrary \textcolor{black}{columns} from $\mathbf{S}$.
		\item  Initialize the resource and the residual matrix $\mathbf{L}$ and  $\mathbf{M}$.
		\item  \textbf{while} new VN $G^v$ comes \textbf{do} 
			\item  \hspace*{0.02in}  \hspace*{0.02in} Build a temporary set $\mathcal{Y}_i=\{\ \}$ to handle conflicts.
			\item  \hspace*{0.02in} \hspace*{0.02in}  \textbf{for} each VM $\mathcal{V}_i$ in VN $G^v$ \textbf{do}
			\item  \hspace*{0.02in}  \hspace*{0.02in} \hspace*{0.02in} \hspace*{0.02in}  Classify $\mathbf{\tilde{\mathcal{D}}}_i$ into pattern $x \in \mathcal{P}$ using model $\emph{m}^*$.	
			\item  \hspace*{0.02in}  \hspace*{0.02in}  \hspace*{0.02in} \hspace*{0.02in} {Select} the $x$-th row of $\mathbf{M}$ as candidate set $F$. 
			\item  \hspace*{0.02in}  \hspace*{0.02in} \hspace*{0.02in}  \hspace*{0.02in} Mute conflicts nodes by setting $F_\rho = 0, \forall \rho \in \mathcal{Y}_i$.
			\item  \hspace*{0.02in}  \hspace*{0.02in} \hspace*{0.02in} \hspace*{0.02in}   \textbf{if} $\nexists F_j \in F, F_j >0$ and $|F|=|\mathcal{N}|$ \textbf{then }
			\item  \hspace*{0.02in}  \hspace*{0.02in} \hspace*{0.02in} \hspace*{0.02in}  \hspace*{0.02in} \hspace*{0.02in}  {Reject} $G^v$ and roll back $\mathbf{L}$ and $\mathbf{M}$; \textbf{break}.
			\item  \hspace*{0.02in}  \hspace*{0.02in} \hspace*{0.02in} \hspace*{0.02in}   \textbf{if} $\exists F_j \in F, F_j >0$ \textbf{then }
			\item  \hspace*{0.02in}  \hspace*{0.02in} \hspace*{0.02in} \hspace*{0.02in}   \hspace*{0.02in} \hspace*{0.02in} {Assign} VM $\mathcal{V}_i$ to server $\mathcal{N}_j$ and add $\mathcal{N}_j$ to $\mathcal{Y}_i$. 
			\item  \hspace*{0.02in}  \hspace*{0.02in} \hspace*{0.02in} \hspace*{0.02in}   \hspace*{0.02in} \hspace*{0.02in} {Update} the residual matrix $\mathbf{M}$. 
			\item  \hspace*{0.02in}  \hspace*{0.02in} \hspace*{0.02in} \hspace*{0.02in}   \textbf{else}
			\item  \hspace*{0.02in}  \hspace*{0.02in} \hspace*{0.02in} \hspace*{0.02in}  \hspace*{0.02in} \hspace*{0.02in}
			\textcolor{black}{{Select} a scheme $\mathcal{S}_y \in \mathcal{S}$ containing $x$.} 
			\item  \hspace*{0.02in}  \hspace*{0.02in} \hspace*{0.02in} \hspace*{0.02in}   \hspace*{0.02in} \hspace*{0.02in} {Label} a new server $\mathcal{N}_{new}$ with scheme $\mathcal{S}_y$.
			\item  \hspace*{0.02in}  \hspace*{0.02in} \hspace*{0.02in} \hspace*{0.02in}   \hspace*{0.02in} \hspace*{0.02in} {Assign} $\mathcal{V}_i$ to server $\mathcal{N}_{new}$ and add $\mathcal{N}_{new}$ to $\mathcal{Y}_i$. 
			\item  \hspace*{0.02in}  \hspace*{0.02in} \hspace*{0.02in} \hspace*{0.02in}   \hspace*{0.02in} \hspace*{0.02in} {Update} the matrix $\mathbf{L}$ and $\mathbf{M}$ using  $\mathcal{S}_y$. 
			\item  \hspace*{0.02in} \hspace*{0.02in}  {Map} virtual nodes in $G^v$ to servers in $\mathcal{Y}_i$ respectively.
			\item  \hspace*{0.02in} \hspace*{0.02in} {Map} virtual links in $G^v$ using shortest path algorithm.
	\end{algorithmic}
\end{algorithm}
For clarity, we group the vectors in $\mathcal{S}$ to form a scheme matrix $\mathbf{S}$, and each column in $\mathbf{S}$ corresponds to a scheme. During the online embedding, we label each active server with a scheme and update its residual vector. The residual vector is initialized to be the same as the scheme vector. A VM with pattern $\mathcal{P}_h$ can only be assign to servers whose $h$-th value of the residual vector is larger than zero. When a VM of pattern $\mathcal{P}_h$ is embedded into a server, the $h$-th value of its residual vector is subtracted by 1. 
The scheme vectors and the residual vectors of servers are collected into $\mathbf{L}$ and $\mathbf{M}$ respectively,  and both matrixs will be updated simultaneously. The new coming VNRs will retrieve available resources from matrix  $\mathbf{M}$.

Initially, we label a small set of servers with arbitrary schemes, and update  $\mathbf{L}$ and $\mathbf{M}$ accordingly.
When a VNR arrives, we quantify each VM $\mathcal{V}_i$ in it to obtain its corresponding pattern $x$.
The $x$-th row of $\mathbf{M}$ will be checked to determine whether there are available resources for pattern $x$.
If there exists one element, e.g., the $y$-th element, in the $x$-th row is greater than 0, the VM will be mapped onto the $y$-th server. Otherwise, a new server will be labeled with a scheme containing pattern $x$ and used to place VM $\mathcal{V}_i$. Only when all the servers run out of resources for patten $x$, will the VNR be rejected.
Moreover, a temporary set can be maintained to avoid conflict, when mutual exclusion constraint is required.
The detailed procedure is specified in Algorithm 3. 

To map a virtual node, only a sparse vector with at most $|\mathcal{N}|$ non-negative integer values is checked. Thus, the worst case complexity of \textbf{Algorithm 3} for a VNR is $\mathcal{O}(|\mathcal{N}| \cdot |\mathcal{N}_v|)$.
In addition, the substrate nodes that can no longer host some patterns can be muted in advance to reduce the complexity. Compared with the traditional online $(d+1)$-dimensional mapping methods, whose computational complexity is
$\mathcal{O}(|\mathcal{N}|^{d+1} \cdot |\mathcal{N}_v|)$,  the performance is improved by at least $|\mathcal{N}|^{d}$.


\section{Testbed Evaluation}


\subsection{Testbed Environment}

Our testbed has 106 servers, connected via SDN-enabled switches via a spine-leaf topology. Each server is equipped with 64-vCPUs, 256G memory, forming a cluster orchestrated by Kubernetes. The transmission rate between any pair of servers is 10Gbps.
Three servers are configured as master,
 which are cordoned off and not allowed to host VMs. 
We define a JSON format to describe each VNR, and implement the proposed algorithm using python language.
The output of the algorithm is in the form of a YAML file and sent to Kubernetes for execution.
In the process pattern combination filtering, the column generation procedure is implemented with the help of both the simplex method in CVX \cite{grant2014cvx} and the Branch and Cut algorithm in Gurobi \cite{gurobi}.


\subsection{Trimmed Alibaba Production Cluster Trace Dataset}

We use the Alibaba production cluster trace dataset \cite{guo2019limits}, which includes 4030 servers and 71,476 VMs. From this, we extract a subset of 142 servers with only online services, resulting in 753 applications and 1,698 VMs. Each VM has CPU and memory demands, and its bandwidth demand is defined as the maximum traffic rate during its lifetime. After normalization, we set the quantization threshold to 1\% for each resource type, yielding 49 patterns. VNRs are then generated based on these patterns for evaluation.

\subsection{Performance Comparison}
Similar with work \cite{naori2020online}, we compared the proposed algorithms with five classical online methods, i.e., \textcolor{black}{First Fit algorithm, Load Balance algorithm, Best Fit algorithm, Kubernetes Default algorithm and Random algorithm} respectively, and carried out \textcolor{black}{20} comparative experiments using stressful workloads on the testbed. For each time, we clean up all the VMs in the 103 servers before we run an algorithm. 
The VNRs used in each comparison experiment for different algorithms are the same.
The number of VMs in each VNR is randomly set according to the distribution of VM number of applications in the dataset. We averaged over \textcolor{black}{20} runs.
\begin{figure}[t!]
	\begin{center}
		\includegraphics[width=88mm]{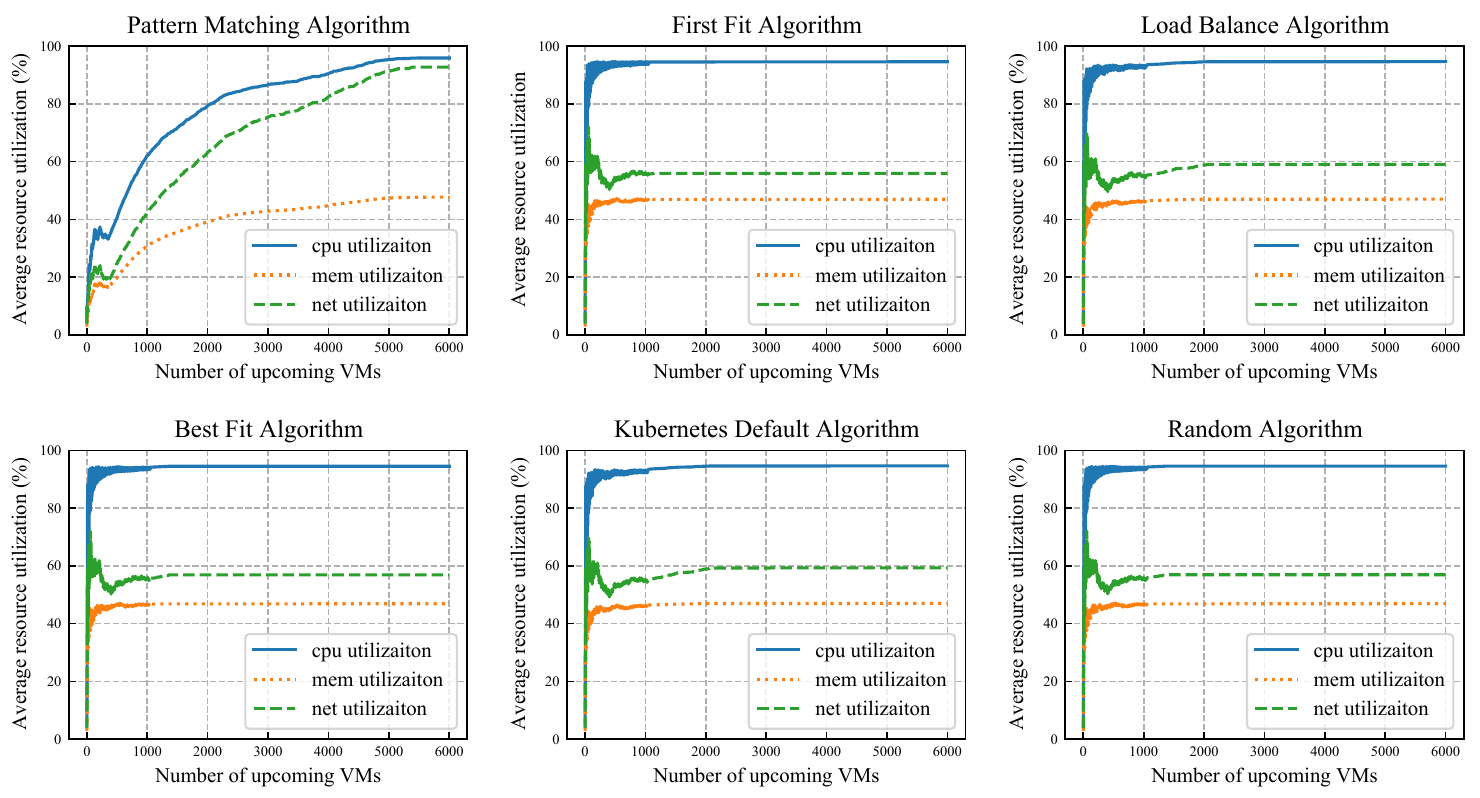}\\
	\vspace{-3 mm}
\caption{Comparison of different algorithms in terms of utilization.}
	\end{center}
\end{figure}

In Fig.~1, we plot the resource utilization curve versus the arrival of VNRs for six different algorithms. It can be seen that \textcolor{black}{the pattern matching algorithm may reject a VM even though it fits one server for further performance}, thus the VM can achieve higher and more balanced long-term resource utilization. On the contrary, for the other five algorithms, the utilization of a CPU resource grows very fast, resulting in the fragmentation and under-utilization of other resources.

In Fig.~2, we compare the average number of maximum VMs accepted by different algorithms. Compared with the other five algorithms, the proposed algorithm can hold 25\%-30\% more VMs, which shows that taking advantage of the admission control mechanism and the complementary relationship can significantly improve the overall performance.

\begin{figure}[t!]
	\centering	\includegraphics[width=66mm]{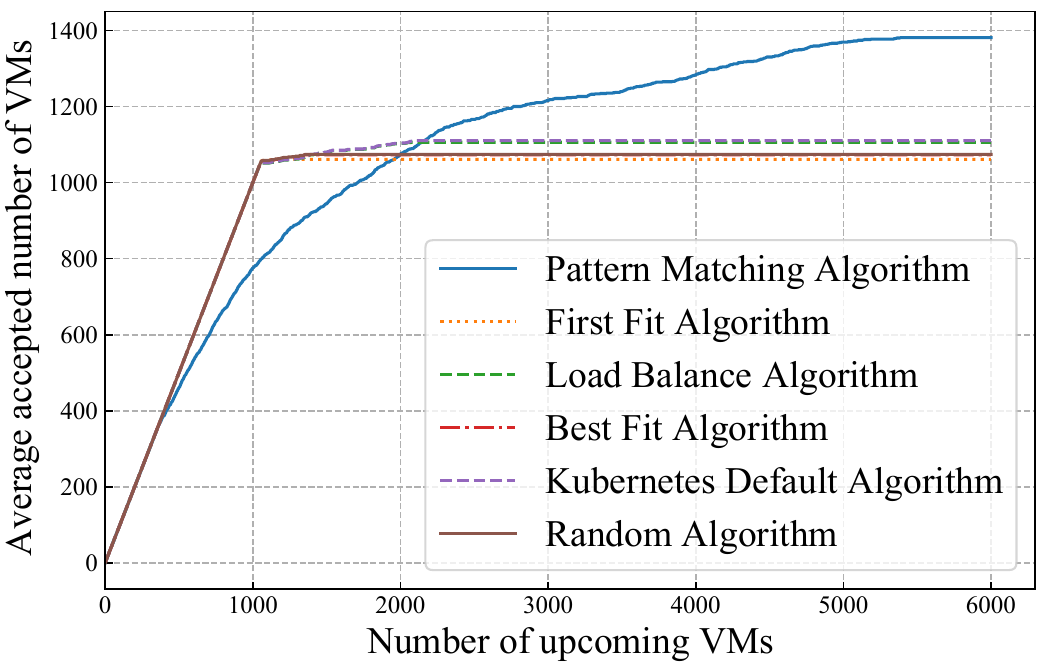}\\
    \vspace{-3 mm}
	\caption{Comparison of different algorithms in terms of the accepted VMs. 
	}
\end{figure}


\section{Conclusion}
In this paper, we have proposed a pattern matching based online VNE approach for cloud data center networks, which exploits the complementary relationship of VNRs to maximize resource utilization. Testbed evaluation using Alibaba production cluster trace dataset has shown that our algorithm achieves close-to-offline performance and more accepted workloads that outperforms traditional designs by 25\%-30\%.
In future, we will develop more dynamic mechanisms and adjustment features to adapt the production environment.

%

\bibliographystyle{IEEEtran}
\bibliography{reference}

\end{document}